\documentclass[12pt]{article}

\usepackage{scicite}
\usepackage{graphicx}
\usepackage[font=footnotesize,labelfont=bf]{caption}
\usepackage{times}

\newenvironment{sciabstract}{%
\begin{quote} \bf}
{\end{quote}}

\title{Nonlinear Schrödinger Kernel for hardware acceleration in machine learning inference}

\author
{Tingyi Zhou,$^{1}$ Fabien Scalzo,$^{1,4,5}$ Bahram Jalali$^{1,2,3\ast}$\\
\\
\normalsize{$^{1}$Department of Electrical and Computer Engineering, UCLA; 420 Westwood Plaza, Los Angeles, 90095}\\
\normalsize{$^{2}$Department of Biomedical Engineering, UCLA; 420 Westwood Plaza, Los Angeles, 90095}\\
\normalsize{$^{3}$California NanoSystems Institute (CNSI); 570 Westwood Plaza, Los Angeles, 90095}\\
\normalsize{$^{4}$Department of Neurology, UCLA; 635 Charles E Young Dr S, Los Angeles, CA 90095}\\
\normalsize{$^{5}$Department of Computer Science, UCLA; 404 Westwood Plaza, Los Angeles, 90095}\\
\\
\normalsize{$^\ast$To whom correspondence should be addressed; E-mail:  jalali@ucla.edu}
}

\date{}

\begin{document} 


\baselineskip24pt


\maketitle


\begin{sciabstract}
  Alternative machine learning approaches that are computationally light with low latency and can work with only a small training dataset are needed for applications where the insatiable demand of deep learning methods for computing power and large training data cannot be met. We show that spectral mapping of data onto femtosecond optical pulses and a projection into an implicit, higher dimensional space using nonlinear optical dynamics increases the accuracy and reduces the latency in data classification by several orders of magnitude. The approach is validated by the classification of various datasets, including brain intracranial pressure, cancer cell imaging, spoken digit recognition, and the classic exclusive OR benchmark for nonlinear classification. The concept is demonstrated by seeding the nonlinear dynamics that are responsible for many fascinating natural phenomena, such as optical rogue waves, with the data before processing the output with a light classifier. A quantitative comparison with a well-known numerical technique is used to provide insight into this physical technique. Single-shot operation is demonstrated using time stretch data acquisition.
\end{sciabstract}


\section*{Introduction}

Neural networks have emerged as a universal computational framework for achieving spectacular performance in image and speech recognition and synthesis. While revolutionary, these performance gains are not without cost. First, the exponential growth in the required computing power outpaces the semiconductor roadmap known as Moore’s law by several orders of magnitude\cite{Xu2018}\cite{thompson2020computational}. Second, the large, labeled datasets required for training these models are not available in many application domains. Third, the latency inherent in deep neural networks can be problematic in real-time applications.

These trends call for a fresh look into the design of machine learning and computing systems that can operate with small training datasets, offer low latency, and do not rely on ever-increasing computational performance and memory size. Among several alternative approaches to computing is the use of analog systems to efficiently perform specialized computations, an approach that echoes the early days of computing \cite{Isaacson2014}. An attractive attribute of analog systems is that they do not need to be trained; however, purely analog computing systems are susceptible to noise and do not scale. A modern approach to analog computing is one where analog does not replace but rather serves as a hardware accelerator for a digital computer \cite{Solli2015}. For example, one can simulate the properties of an inaccessible and complex system, such as that underlying hydrodynamic phenomena, with a more user-friendly proxy system, such as wave propagation in a nonlinear optical fiber, which is governed by the same differential equations. A natural computer based on such rapid dynamics can serve as a surrogate for the computation of fluid dynamics phenomena \cite{Solli2015}. When considering the vast parameter space that needs to be explored, the benefits are even greater. A related framework could involve the use of wave propagation in a metamaterial to perform specialized computational tasks such as solving specific integral equations \cite{Estakhri2019}. With an instrument capable of capturing the output of such natural computers in real time, billions of scenarios can be readily acquired on an ultrashort timescale to map the vast space of complex outputs that emerge from the nonlinear responses to different inputs. Another example of physical computing pertains to the emulation of a neural network with optical circuits. The goal here is to use lasers and spatial light modulators or integrated optical circuits to mimic the architecture of a neural network. Specific recent examples include the use of spatial light modulators for matrix manipulation or using integrated optical circuits \cite{Lin2018}\cite{George2019}\cite{wetzstein2020inference}. Neural networks implemented in software have been utilized for solving inverse problems in optics, including phase retrieval for ultrashort pulse reconstruction using frequency resolved optical gating (FROG) \cite{krumbugel1996direct}\cite{zahavy2018deep} and the analysis of modulation instability in an optical fiber \cite{narhi2018machine}.

In this paper, we report a new concept in data representation and classification. Specifically, we show that when seeded with data, nonlinear optical dynamics enable a linear learning algorithm to learn a nonlinear function or decision boundary that separates the data into the correct classes. The core of this system, which we call the Nonlinear Schrödinger Kernel, nonlinearly projects data that have been modulated onto the spectrum of a femtosecond pulse into a space in which the data, which normally could not be linearly separated, can now be classified with a linear classifier.

Figure 1 shows a qualitative interpretation of the working principle of this model. The input data are first mapped to virtual spectrum nodes by a spectral modulator and then undergo a nonlinear transformation—described by the Nonlinear Schrödinger Equation (NLSE)—causing complex energy and phase transfer between the nodes. The experimental system is described later.

As a proof of concept for our model, we first demonstrate the exclusive OR (XOR) operation, a classic nonlinear classification benchmark. We then obtain classification results on a variety of data, including brain intracranial pressure (ICP), spoken digit, and biological cell data acquired with time stretch imaging. We show that the combination of the Nonlinear Schrödinger Kernel with a digital classifier provides improvements in accuracy and latency over a conventional, purely digital classifier. To provide insight into the operation of the Nonlinear Schrödinger Kernel, we compare its performance to the radial basis function (RBF), the most common kernel in machine learning, and show that both enable a linear classifier to perform nonlinear classification. Finally, the operation of the Nonlinear Schrödinger Kernel is demonstrated with a wide variety of linear and nonlinear digital algorithms, and its robustness against quantization and additive noise is illustrated.

\begin{figure}
\centering
\includegraphics[width=\linewidth]{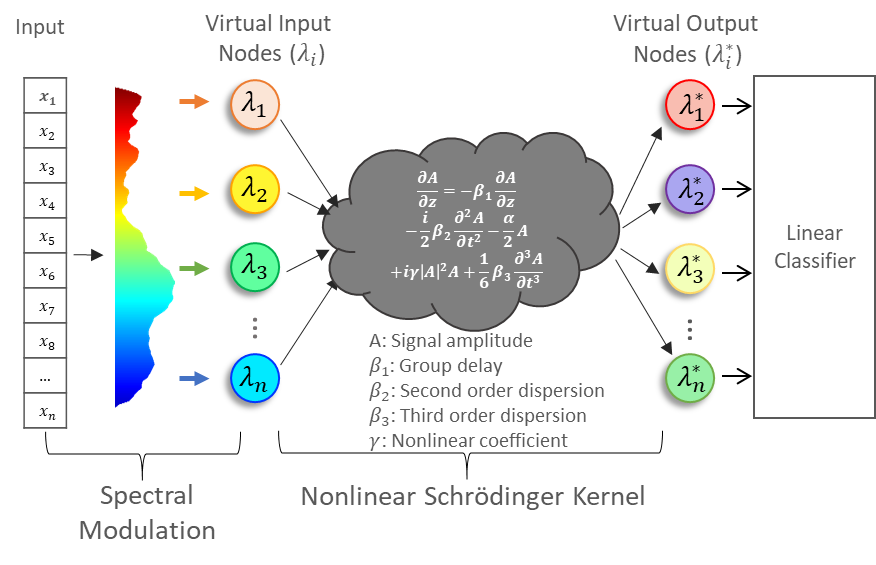}
\caption{The Nonlinear Schrödinger Kernel computing framework. Samples of the input data are first assigned to virtual nodes through spectral modulation. The input data are nonlinearly projected to an output representation with a transformation described by the Nonlinear Schrödinger Equation (NLSE). The experimental system is described in the paper}
\label{fig:fig1}
\end{figure}

Since this technique operates with femtosecond pulses and spectrally modulated data, it is inherently compatible with time stretch data acquisition, including time stretch microscopy and spectroscopy, for continuous single-shot acquisition and classification of ultrafast events \cite{Kelkar1999}. To this end, successful operation with both a conventional grating spectrometer and the time stretch technique is demonstrated. Spectral mapping and electric-field transformation, governed by the Nonlinear Schrödinger Equation, enable certain nonlinear classification problems to be solved with a low-latency linear classifier by transforming the data into an implicit high-dimensional representation.

\subsection*{Demonstrated tasks}

We illustrate the effectiveness and versatility of the proposed technique on several classification tasks, including XOR, the detection of intracranial hypertension \cite{scalzo2012reducing} time stretch microscopy cell image classification for cancer detection \cite{Chen2016}, and recognition of spoken digits \cite{zohar_jackson_2018_1342401}. All the datasets include ground-truth class labels that are used to train a machine learning model in a supervised manner before and after processing by the Nonlinear Schrödinger Kernel. 

\subsection*{Experiment description}

As shown in figure 2a, the Nonlinear Schrödinger Kernel computing system consists of a fiber-based supercontinuum laser source, a spectral modulator, a nonlinear optical element, and a spectrum readout followed by a machine learning classifier operating in the digital domain. The data are first modulated onto the laser spectrum through the spectral modulator. The modulated waveform then passes through a nonlinear optical element where the data are nonlinearly transformed. We define the combination of the spectral modulation and the nonlinear transformation as the Nonlinear Schrödinger Kernel. Because of the spectral nature of this kernel, it can also be called the Lambda Kernel, in homage to the convention of representing wavelengths with the symbol lambda ($\lambda$). The optical spectrum output from the Nonlinear Schrödinger Kernel is read out and fed into the digital signal processor (DSP) for machine learning classification. The nonlinear element in the experiment is a highly nonlinear fiber; however, via simulation, we verified that our technique will also work with silicon- or silicon nitride-integrated waveguides (Table S1). The spectrum can be readout by either a traditional grating-based spectrometer or time stretch spectrometer for fast single-shot operation \cite{Kelkar1999}. Techniques with both devices are demonstrated in this work.

\subsection*{Machine learning models}

As part of our experiments, we evaluate the performance of popular machine learning algorithms with and without preprocessing using the Nonlinear Schrödinger Kernel. These algorithms include ridge regression \cite{hoerl1970ridge}, decision trees \cite{ho1995random}, spectral regression for kernel discriminant analysis (SR-KDA) \cite{cai2011speed}, neural networks, and support vector machines (SVMs) \cite{cortes1995support}. For all these methods, it is assumed that the input data are sampled as $X\in R^{n\times d}$   and the output class label as $Y\in R^{n\times c}$, where $c$ is the number of classes. The dimensionality $d$ and the number of samples $n$ of the input data vary with respect to the dataset used. 
\begin{figure}
\centering
\includegraphics[width=0.8\linewidth]{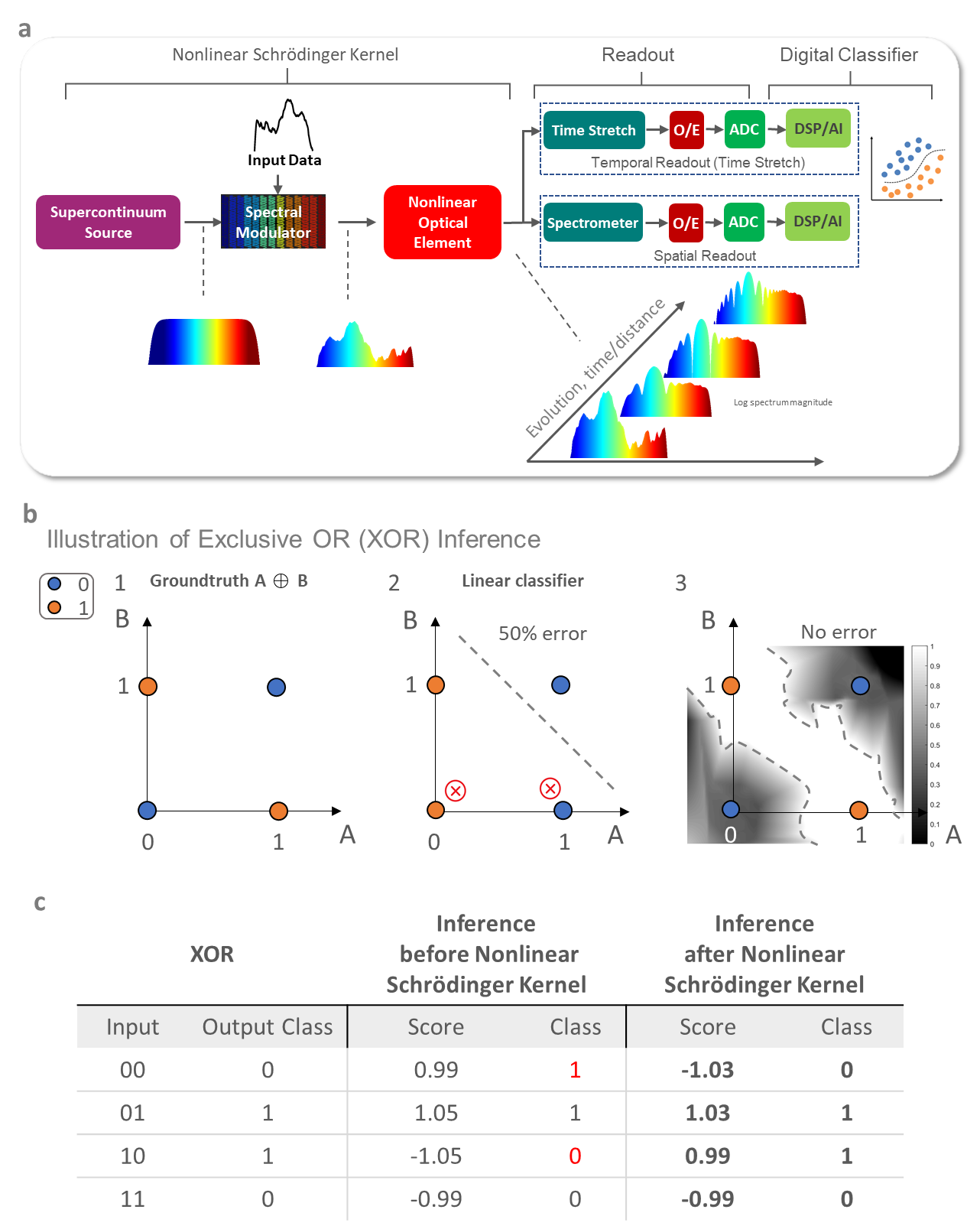}
\caption{Block diagram of the experimental setup (a) and the classification results for the Exclusive OR (XOR) operation (b and c). The XOR operation is the most widely used benchmark for demonstrating nonlinear classification. (a) Simplified block diagram showing the basic building blocks. A broadband femtosecond mode-locked laser is used as the spectral source. The data are modulated onto the optical spectrum and transformed by nonlinear propagation in an optical medium. The inset shows the evolution of the data-modulated spectrum. The spectrum at the output is read by a conventional spectrometer based on diffraction gratings or by time stretch spectroscopy for fast single-shot performance (examples of each are shown). Classification is performed using a machine learning algorithm in the digital domain.\\
(b) Demonstration of nonlinear classification using a linear classifier. The data points are not linearly separable, but the Nonlinear Schrödinger Kernel maps them into a space where they are. 2b.1 shows the XOR ground truth. 2b.2 shows that a linear classifier cannot produce the correct output. 2b.3 is the output of a linear classifier preceded by the Nonlinear Schrödinger Kernel, showing an error-free XOR operation. The class probability predicted by the linear classifier after model training on 4 pairs of inputs is mapped in grayscale over a range from 0 (black) to 1 (white). The boundary of the classifier output is outlined with dashed gray lines, which can be seen to separate the expected output values of the 4 possible pairs. \\
(c) The table shows the scores and predicted classes output by a linear support vector machine (SVM) classifier before and after processing with the Nonlinear Schrödinger Kernel. These results show that without the Nonlinear Schrödinger Kernel, the classification fails, while implementation of the Kernel results in successful classification. The scores show that the accuracy improvement is caused by the projection of the data into a space in which the scores are transformed into a linearly separable array.
}
\label{fig:fig2}
\end{figure}

\subsection*{Performance evaluation}

The performance of each machine learning model is evaluated using 3-fold cross-validation, and the accuracy is reported using the area under the receiver operating characteristic (ROC) curve (AUC). The AUC represents the probability that the model will be able to predict the label correctly for a new pair of positive and negative samples. With the same cross-validation split, we compare the AUC of each model with the original data as input before and after processing with the Nonlinear Schrödinger Kernel, and we repeat the experiment protocol on each dataset.

\section*{Results}

The effectiveness of the proposed approach is evaluated through a number of experimental protocols. An interesting property of the Nonlinear Schrödinger Kernel is that it projects the original signal into a modified space that emphasizes certain nonlinear properties of the data. In terms of functionality, similarities exist between the Nonlinear Schrödinger Kernel and the concept of “kernel projection” or “the kernel trick” in the machine learning literature. The utility of this processing method is that it transforms nonlinearly separable data to become linearly separable in the new, modified space.

We first illustrate this property by performing the XOR task, a classic problem that cannot be solved with a linear model and often serves as a benchmark for nonlinear classification. When four points representing the binary input pairs ([0,0], [0,1], [1,0], [1,1]) and their corresponding XOR output (0,1,1,0) are used as the training data, linear classifiers will inevitably result in an error when attempting to reproduce the results of the XOR operation (as illustrated in Fig. 2b.2) because the data points are linearly nonseparable in their original input space. On the other hand, after processing the data with the Nonlinear Schrödinger Kernel, a linear classifier can be trained to successfully perform the XOR operation error free (Fig. 2b.3).

To gain insight into how the Nonlinear Schrödinger Kernel projects the input data into a linearly separable representation, the class probability predicted by the linear classifier is represented by a grayscale map with values ranging from 0 (black) to 1 (white), as shown in Fig. 2b.3. The boundaries of the grayscale map are depicted as dashed lines, which clearly show the accurate separation of the four pairs of inputs. This example reveals that with the use of the Nonlinear Schrödinger Kernel, the linearly nonseparable data points are projected to a space where they become linearly separable. An important property of the Nonlinear Schrödinger Kernel projection is that it is unsupervised; in other words, the output labels of the pairs of data points are not used to obtain the projected values in the new space.

For further intuition into the transformation performed by the combined operation of spectral modulation and nonlinear evolution, we examine the scores and the predicted classes obtained by a linear SVM classifier with and without the application of the Nonlinear Schrödinger Kernel (Fig. 2c). The scores, calculated as the inner products of the classifier weights and the input data, show that the accuracy improvement is caused by the modification of the score into a linearly separable form.

Having demonstrated the ability of the Nonlinear Schrödinger Kernel to perform nonlinear projection to enable linear classification, we then test the hypothesis that the kernel may act as a nonlinear feature extractor that highlights certain properties of the signal and makes challenging signal classification problems solvable with a simple machine learning model. In particular, we focus on several challenging tasks with varying degrees of complexity, including the detection of brain intracranial hypertension, classification of blood cell images for cancer detection, and recognition of spoken digits.

\begin{figure}
\centering
\includegraphics[width=\linewidth]{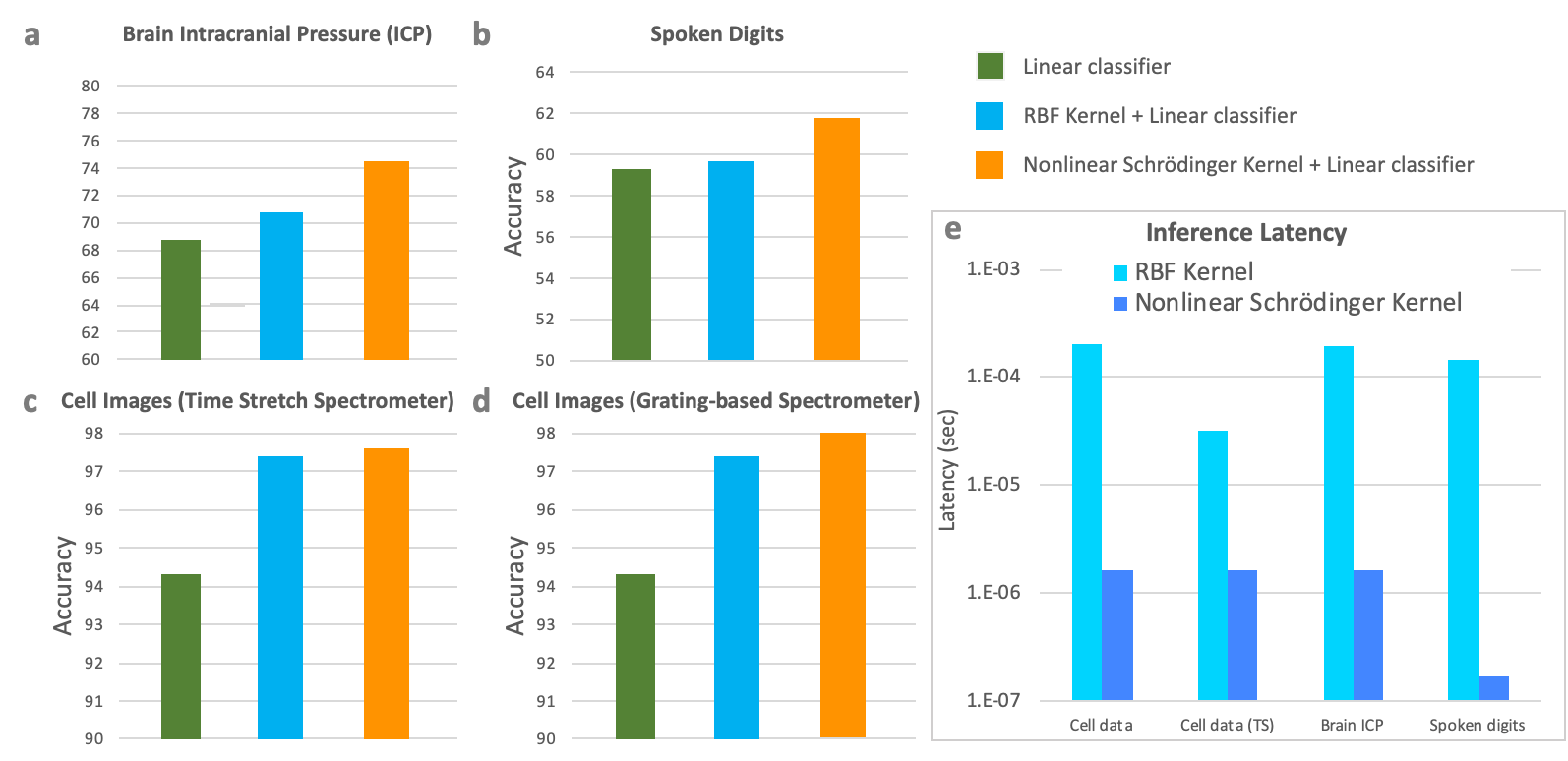}
\caption{Results for four different datasets: (a) brain intracranial pressure (ICP), (b) spoken digits, (c) microscope-derived cancer cell images with a time stretch (TS) spectrometer Kernel readout, and (d) microscope-derived cancer cell images with a grating-based spectrometer Kernel readout. For each dataset, the area under the curve (AUC) is reported after threefold cross-validation. For all four datasets, compared with the linear classifier alone, the (physical) Nonlinear Schrödinger Kernel shows improvement that is similar to or slightly better than that obtained by the (numerical) radial basic function (RBF) kernel. (e) A bar chart shows the latency in the processing of one sample through the RBF kernel and the Nonlinear Schrödinger Kernel. The results are grouped by dataset, including cell data using the grating-based spectrometer readout, cell data using the TS spectrometer readout, ICP, and the audio signals recorded from spoken digits. The bar chart shows a latency (on a Xeon 3 GHz CPU with 64 GB of RAM) on the order of 1E-4 to 1E-5 seconds for the RBF kernel, while the Nonlinear Schrödinger Kernel achieved a substantially reduced latency, on the order of 1E-6 to 1E-7. While a highly nonlinear fiber was used in these experiments, the same 3rd-order optical nonlinearities are also present in integrated waveguides composed of materials such as silicon and silicon nitride. If implemented using such waveguides, the latency of the Nonlinear Schrödinger Kernel will be in the picosecond regime.}
\label{fig:fig3}
\end{figure}

Fig. 3 provides the classification results for four datasets: brain ICP (3a), spoken digit (3b), and cancer cell image data with two different Nonlinear Schrödinger Kernel readouts: one using a time stretch spectrometer (3c) and the other using a grating-based spectrometer (3d). The accuracy is calculated using the AUC after a 3-fold cross-validation. As a benchmark, we compare the performance of the Nonlinear Schrödinger Kernel with the most commonly used numerical kernel, the RBF, for performing nonlinear classification with a linear classifier. When the same data are applied to both models, the Nonlinear Schrödinger Kernel reaches an accuracy similar to or slightly better than the numerical RBF kernel. These results suggest that the physical processes of spectral modulation and nonlinear evolution perform, with much lower latency (Fig. 3e), a conceptually similar task as the so-called “kernel trick” in machine learning literature.

While Fig. 3 highlights the benefit of the Nonlinear Schrödinger Kernel when implemented with a linear classifier, we now turn to more advanced machine learning classifiers that already include nonlinearity in their models, including decision trees, SR-KDA, SVM, ridge regression, and neural network. Specifically, we aim to test whether the Nonlinear Schrödinger Kernel can further improve the result obtained with these already nonlinear classifiers. The bar chart in Fig. 4 summarizes the experimentally measured performance of the Nonlinear Schrödinger Kernel with five different nonlinear classification algorithms. The AUC is reported using the original data as input before (gray) and after processing with the kernel (blue). Consistent improvement in the AUC can be observed across the different classification methods; for example, average $11.6\%$ improvement in AUC was observed for the brain ICP dataset. The table shows that similar improvement is observed for the other evaluated datasets (time stretch microscopy cell image and spoken digit datasets), revealing that the Nonlinear Schrödinger Kernel substantially outperforms the original datasets.

\begin{figure}
\centering
\includegraphics[width=\linewidth]{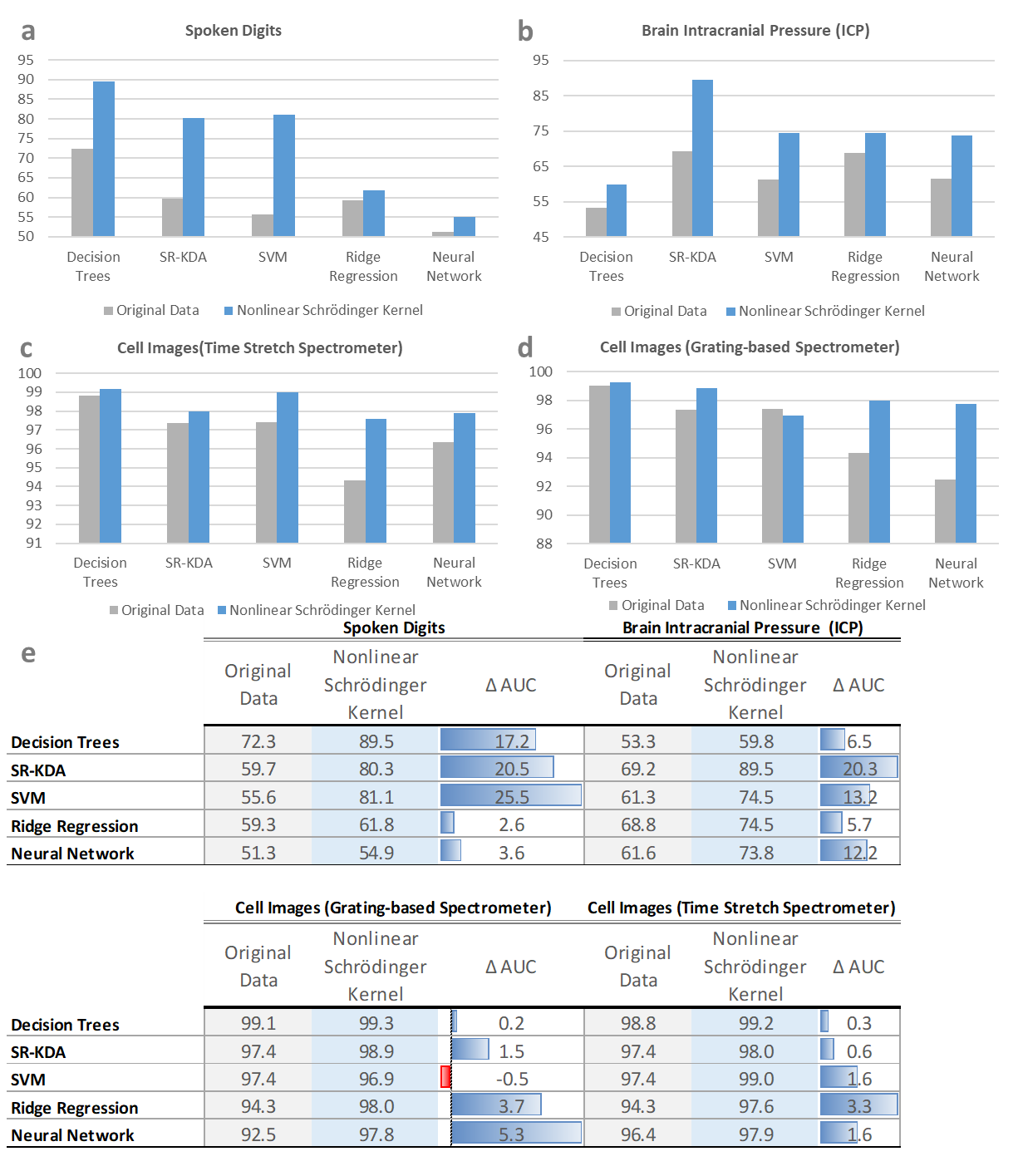}
\caption{Classification results with different machine learning algorithms. The performance of the Nonlinear Schrödinger Kernel in the classification of various datasets (tasks) is shown: spoken digit (a), brain intracranial pressure (ICP) (b), cell image data with time stretch spectrometer readout (c), and cell image data with grating-based spectrometer readout (d). For each dataset, we show the performance with and without the optical Kernel for five different classifiers: decision trees, spectral regression for kernel discriminant analysis (SR-KDA), support vector machine (SVM), ridge regression, and neural network. The area under the curve (AUC) is reported using the original data as input before (gray) and after processing with the Nonlinear Schrödinger Kernel (blue). The improvement achieved following processing with the Nonlinear Schrödinger Kernel is consistent for different algorithms and on all datasets, showing that the Nonlinear Schrödinger Kernel can improve the performance of a wide range of machine learning algorithms, both linear and nonlinear, as demonstrated in the tables of classification accuracies in (e).}
\label{fig:fig4}
\end{figure}

We next investigate the limitations imposed by signal detection. When dealing with physical systems that are inherently analog in nature, the quantization noise of the analog to digital converter, as measured by its number of bits, can act as a performance bottleneck. This is particularly true for real-time systems that operate at high bandwidth \cite{mahjoubfar2017time}. In addition, the number of bits used to store the model weights also needs to be considered because despite the improvement in accuracy, a higher number of bits requires larger memory. To this end, Fig. 5 shows the robustness of the Nonlinear Schrödinger Kernel-projected representations across the various quantization resolutions applied to the input and internal weights of the model. For this purpose, we utilize the SR-KDA model and quantize the model from 6 to 2 bits while simulating white Gaussian noise of 0, 5, and $10\%$ of the data. The noise is added to the data after it is processed with the Nonlinear Schrödinger Kernel, and the error is evaluated for each combination of noise and quantization precision using the relative mean squared error (MSE), which is calculated between the predicted output of the ridge regression model and the ground-truth label and reported for various levels of quantization precision for the model weights. Finally, a polynomial fit is used to plot a continuous line that best fits the error values. For the cell image and spoken digit data, the MSE increases steadily with noise and a decrease in the number of bits. However, for the brain ICP data, quantization does not affect the results in low noise conditions, and the MSE is observed to remain constant. 

\begin{figure}
\centering
\includegraphics[width=\linewidth]{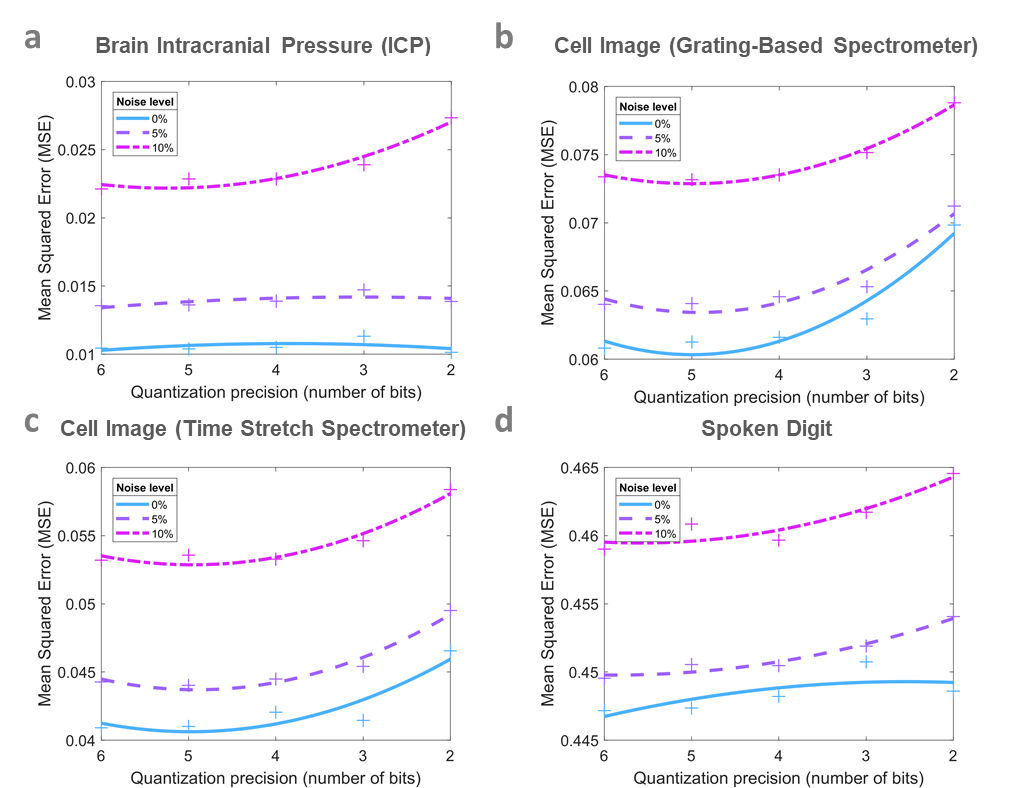}
\caption{The robustness of the Nonlinear Schrödinger Kernel-based model against additive and quantization noises. A spectral regression for kernel discriminant analysis (SR-KDA) model was trained on 4 datasets processed after the data were projected into an implicit high-dimensional space with the Nonlinear Schrödinger Kernel. The mean squared error (MSE) between the predicted output of the model and the ground truth is reported for various levels of quantization precision (from 2 to 6 bits). Three levels of additive noise are reported: no noise, $5\%$ and $10\%$. The lines are polynomial fits to the data points.}
\label{fig:fig5}
\end{figure}

\section*{Discussion}

Recent breakthroughs in machine learning and advanced computing have resulted in numerous applications of data processing and classification. The need for low-latency data processing frameworks that capture nonlinear signal features without requiring any training is particularly important in modern technologies related to self-driving vehicles and healthcare and sensing systems. In this paper, we introduced a data processing framework that utilizes the nonlinear dynamics of an optical system to modulate and process data that are modulated onto the spectrum of the system. When combined with standard machine learning algorithms, this processing framework, named the Nonlinear Schrödinger Kernel, demonstrates significant improvement over existing linear frameworks in classification accuracy on a wide range of applications at a much lower latency, therefore offering a desirable balance between the two.



\bibliography{Lib.bib}

\bibliographystyle{Science}

\section*{Acknowledgments}
This work is supported by the DARPA-MTO PEACH program under contract HR00111990050.



\clearpage

\end{document}